\documentstyle[12pt]{article}

\def\+{{+\!\!\!+}}
\def\pp{\mbox{\tiny${}_{\stackrel\+ =}$}}

\def\ra{\rightarrow}

\def\d{\partial}

\def\th{\theta}

\def\P{\Phi}

\def\p{\psi}

\def\e{\varepsilon}

\def\bR{\hbox{R\hspace{-0.10in}I\hspace{0.04in}}} 
\def\bC{\hbox{C\hspace{-0.10in}I\hspace{0.04in}}} 
\def\pmb#1{\setbox0=\hbox{#1}%
\kern.0em\copy0\kern-\wd0
\kern-.04em\copy0\kern-\wd0
\kern.08em\copy0\kern-\wd0
\kern-.04em\raise.0433em\box0 }         

\def\half{\frac{1}{2}}
\def\rank{\textstyle{\rm{rank}}}

\newcommand{\nc}{\newcommand}
\nc{\beq}{\begin{equation}}
\nc{\eeq}[1]{\label{#1}\end{equation}}
\nc{\ber}{\begin{eqnarray}}
\nc{\eer}[1]{\label{#1}\end{eqnarray}}
\nc{\pek}[1]{\cite{#1}}
\nc{\enr}[1]{(\ref{#1})}
\nc{\kal}[1]{{\cal{#1}}}
\nc{\dott}{\;\cdot\;}
\newcommand{\Section}[1]{\section{#1} \setcounter{equation}{0}}

\begin{document}
\newcommand{\inv}[1]{{#1}^{-1}} 
\renewcommand{\theequation}{\thesection.\arabic{equation}}
\newcommand{\be}{\begin{equation}}
\newcommand{\ee}{\end{equation}}
\newcommand{\bea}{\begin{eqnarray}}
\newcommand{\eea}{\end{eqnarray}}
\newcommand{\re}[1]{(\ref{#1})}
\newcommand{\qv}{\quad ,}
\newcommand{\qp}{\quad .}
\begin{center}
                       
                                \hfill   hep-th/0202069\\
\vskip .3in \noindent

\vskip .1in

{\large \bf{N=1 supersymmetric sigma model with boundaries, II}}
\vskip .2in

{\bf Cecilia Albertsson}$^a$\footnote{e-mail address: cecilia@physto.se},
{\bf Ulf Lindstr\"om}$^a$\footnote{e-mail address: ul@physto.se}
 and  {\bf Maxim Zabzine}$^{b}$\footnote{e-mail address: zabzine@fi.infn.it} \\

\vskip .15in

\vskip .15in
$^a${\em  Institute of Theoretical Physics \\
Stockholm Centre for Physics, Astronomy and Biotechnology \\
University of Stockholm,
SE-106 91 Stockholm, Sweden}\\
\vskip .15in
$^b${\em INFN Sezione di Firenze, Dipartimento di Fisica\\
 Via Sansone 1, I-50019 Sesto F.no (FI), Italy}

\bigskip

\vskip .1in

\end{center}
\vskip .4in

\begin{center} {\bf ABSTRACT } 
\end{center}
\begin{quotation}\noindent 
We consider the $N=1$ supersymmetric two-dimensional non-linear sigma
model with boundaries and nonzero $B$-field.  By analysing the
appropriate currents we describe the full set of boundary conditions
compatible with $N=1$ superconformal symmetry.  Using this result the
problem of finding a correct action is discussed. We interpret the
supersymmetric boundary conditions as a maximal integral submanifold
of the target space manifold, and speculate about a
new geometrical structure, the deformation of an almost product
structure.
\end{quotation}
\vfill
\eject

\Section{Introduction}

The conditions that arise from the two-dimensional non-linear sigma
model when imposing $N=1$ worldsheet supersymmetry on the boundary
have some interesting implications.\footnote{For a related discussion
treating BRST invariance see \cite{LRvN}.} These boundary conditions
may be interpreted in terms of the target space manifold, where they
put restrictions on the way in which D-branes may be embedded.

The present paper is a continuation and generalisation of the results
presented in our previous paper \cite{Albertsson:2001dv}.  There we
considered a background of general metric and zero $B$-field, and
showed how worldsheet supersymmetry on the boundary leads to the
appearance of Riemannian submanifolds of the target space ${\cal M}$.
Here we extend the analysis to a background $E_{\mu\nu}= g_{\mu\nu} +
B_{\mu\nu}$, where both the metric $g_{\mu\nu}$ and the antisymmetric
field $B_{\mu\nu}$ are general.  It turns out that Riemannian
submanifolds arise here again, but in a slightly different way.  In a
sense, the $B\neq 0$ case is a deformation of the $B=0$ case.

The bulk dynamics of the non-linear sigma model is given by the
superfield action (for definitions, see Appendix~\ref{a:susy})
\beq
 S= \int d^2\sigma d^2\theta \,\,D_{+} \Phi^\mu D_- \Phi^\nu E_{\mu\nu}
(\Phi).
\eeq{action1}
The conserved currents can be consistently derived from
this action. However, we do not know a priori whether or not
(\ref{action1}) is the correct action for deriving the boundary
dynamics. Therefore our starting point is to derive the
boundary conditions by looking at the appropriate boundary conditions
for the bulk conserved currents,
\beq
 \left[ \,\, T_{++}-T_{--} \,\,
 \right]_{\sigma=0,\pi} =0,\,\,\,\,\,\,\,\,\,\,\,\,\,\,\,
 \left[ \,\,  G_{+}-\eta G_{-} \,\, \right]_{\sigma=0,\pi} =0 ,
\eeq{bulkcurbc}
where $T_{\pm\pm}$ and $G_{\pm}$ correspond to, respectively,
the stress tensor and the supersymmetric current.
We then define the correct action to be such that
it reproduces the full set of conditions resulting from
(\ref{bulkcurbc}). We show that this action can be obtained by
adding a unique boundary term to (\ref{action1}).  The
present discussion serves as a clarification of the
work \cite{Haggi-Mani:2000uc} where the problem with the standard
action was pointed out.

This paper is organised as follows. In Section~\ref{s:axioms} we
introduce and motivate the set of axioms that serve as our starting
point for the analysis. In Section~\ref{s:currents} we derive the
boundary conditions that follow from $N=1$ superconformal invariance
at the level of the conserved currents. Section~\ref{s:action1}
discusses the correct action in elementary terms, while
Section~\ref{s:action2} provides the general derivation of this
action. In Section~\ref{s:geometry} we interpret the supersymmetric
boundary conditions as a maximal integral submanifold of the target
space, and we also study the effect the $B$-field has on the boundary
conditions. In Section~\ref{s:global} we speculate about the possible
global interpretation of the boundary conditions in the presence of a
$B$-field.  In our view this leads to an interesting geometrical
structure which we call a deformation of an almost product structure.
Finally, in Section~\ref{s:discussion} we discuss some possible
directions for future work and the relation of our results to those
usually adopted in the literature.

\Section{Algebraic considerations}
\label{s:axioms}

We begin by setting the stage for our investigation, by
adopting some definitions. To motivate the first assumption that we will
make, let us first recall some definitions adopted in our analysis
of the $B=0$ case \cite{Albertsson:2001dv}.
There we introduced two
orthogonal projectors, $P^\mu_{\,\,\nu}$ and $Q^\mu_{\,\,\nu}$, which
were written in terms of a (1,1) ``tensor'' $R^\mu_{\,\,\nu}$ that
represented the worldsheet boundary
conditions\footnote{$R^\mu_{\,\,\nu}$ is not a proper tensor field
since it is not required to be defined on the entire spacetime manifold.}
($\eta \equiv \pm 1$),
\beq
 \left[ \,\, \psi_-^\mu- \eta R^\mu_{\,\,\nu} \psi_+^\nu \,\,
\right]_{\sigma=0,\pi} =0,
\eeq{alferm}
and which squared to one, $R^\mu_{\,\,\nu} R^\nu_{\,\,\rho}=
\delta^\mu_{\,\,\rho}$. (From now on we shall drop the notation $[
\cdot\cdot\cdot ]_{\sigma=0,\pi}$, since all conditions in this paper are
understood to hold on the boundary.)  $P^\mu_{\,\,\nu}$ and
$Q^\mu_{\,\,\nu}$ could be thought of as projectors onto the Neumann
and Dirichlet directions, respectively, so that, e.g., the
covariant Dirichlet condition is written $Q^\mu_{\,\,\nu} \d_0 X^\nu =
0$.  Moreover, we introduced so-called \emph{adapted coordinates}, a
basis in which $P$, $Q$ and $R$ take the form
\beq
P^\mu_{\,\,\,\nu} = \left ( \begin{array}{cc}
            \delta^n_{\,\,\,m} & 0 \\
              0 & 0
           \end{array} \right ) ,
\,\,\,\,\,\,\,
Q^\mu_{\,\,\,\nu} = \left ( \begin{array}{cc}
              0 & 0 \\
              0 & \delta^i_{\,\,\,j}
           \end{array} \right ) ,
\,\,\,\,\,\,\,
R^\mu_{\,\,\,\nu} = \left ( \begin{array}{cc}
            \delta^n_{\,\,\,m} & 0 \\
              0 & -\delta^i_{\,\,\,j}
           \end{array} \right ) ,
\eeq{e:PQRold}
where $n$ and $m$ are spacetime indices in the Neumann directions,
$i$ and $j$ are Dirichlet indices, and $\delta^\mu_{\,\,\,\nu}$
is the Kronecker delta.
In particular, the Dirichlet boundary conditions in this
basis assume their familiar form,
\beq
\psi_-^i + \eta \,\psi_{+}^i =0.
\eeq{e:Dirichletbc}

Our starting point in the present study is the
Dirichlet projector $Q^\mu_{\,\,\nu}$. The Dirichlet conditions
are by definition the same as for $B=0$, so we may
again assume the existence of a projector $Q$ such that
\beq
Q^\mu_{\,\,\nu} \d_0 X^\nu = 0 ,
\,\,\,\,\,\,\,\,\,\,\,\,
Q^\mu_{\,\,\rho} Q^\rho_{\,\,\nu} = Q^\mu_{\,\,\nu} .
\eeq{QQ}
In addition, we make the same ansatz (\ref{alferm}) for the fermionic boundary
conditions as we did for $B=0$, and we require that $R^\mu_{\,\,\nu}$
satisfy
\beq
Q^\mu_{\,\,\rho} R^\rho_{\,\,\nu} = R^\mu_{\,\,\rho} Q^\rho_{\,\,\nu}
= - Q^\mu_{\,\,\nu}.
\eeq{QR}
This property is reasonable from a physical point of view;
in adapted coordinates we write
\beq
Q^\mu_{\,\,\nu}=\left(
\begin{array}{cc}
0&0 \\
0&\delta^i_{\,\,j}
\end{array} \right) ,
\,\,\,\,\,\,\,\,\,\,\,
R^\mu_{\,\,\nu}=\left(
\begin{array}{cc}
R^m_{\,\,n}&0 \\
0&-\delta^i_{\,\,j}
\end{array} \right) ,
\eeq{e:adapted}
which clearly satisfy (\ref{QR}).
We had the same property for $B=0$, but note that here
$R^\mu_{\,\,\nu}$ does not square to one.

We now proceed to introduce the object $P^\mu_{\,\,\nu}$,
\beq
P^\mu_{\,\,\nu}= \half (\delta^\mu_{\,\,\nu}+R^\mu_{\,\,\nu}) ,
\eeq{Pdef}
which, unlike in the $B=0$ case, is not a projector.
However,
we may still derive some important results
in the same way as in \cite{Albertsson:2001dv},
as follows.
Applying the off-shell supersymmetry transformations (\ref{compsusytr})
to the ansatz (\ref{alferm}), we
obtain
\beq
\d_{=} X^\mu - R^\mu_{\,\,\nu}\d_{\+}X^\nu - 2i\eta P^\mu_{\,\,\nu}
F_{+-}^\nu + 2iR^\mu_{\,\,\nu,\sigma} P^\sigma_{\,\,\rho}
\psi^\rho_{+} \psi^\nu_{+} =0 .
\eeq{bbcNt}
Inserting the $F$-field equations
\beq
F^\rho_{+-} + \Gamma^{-\rho}_{\,\,\lambda\sigma}
\psi_{+}^\lambda \psi_{-}^\sigma =0,
\eeq{Feom}
where $\Gamma^{-\rho}_{\,\,\lambda\sigma}$ is the connection
with torsion defined in Appendix~\ref{a:connections}, (\ref{bbcNt})
reduces to its on-shell version,
\beq
\d_= X^\mu - R^\mu_{\,\,\nu}\d_\+ X^\nu + 2i(P^\sigma_{\,\,\rho}
\nabla_\sigma R^\mu_{\,\,\nu} + P^\mu_{\,\,\gamma} g^{\gamma\delta}
H_{\delta\sigma\rho} R^\sigma_{\,\,\nu})\psi_+^\rho \psi_+^\nu =0 .
\eeq{susybosRRR}
Here $H_{\mu\nu\rho}$ is the field strength of the background $B$-field.
We next contract (\ref{susybosRRR}) with $Q^\lambda_{\,\,\mu}$,
and use that $Q^\mu_{\,\,\nu}\d_0 X^\nu =0$, to
obtain\footnote{Note that the property (\ref{QR}) is crucial to finding
(\ref{dbzeroinegr}). Conversely, one might take the view that
contracting (\ref{susybosRRR}) with $Q$ should lead to
$Q^\mu_{\,\,\nu} \d_0 X^\nu =0$ in the same way as it did for
$B=0$. To achieve this, (\ref{QR}) is required, and thus it is
implied by our assumption of Dirichlet conditions.}
\beq
P^\mu_{\,\,\nu} P^\rho_{\,\,\sigma} \nabla_{[\rho} Q^\delta_{\,\,\mu]} = 0.
\eeq{dbzeroinegr}
This equation looks exactly like the condition we
found in the $B=0$ case, and which led to the integrability condition
for $P$. However, as $P$ is not a projector, this
interpretation does not apply here.

To find out what (\ref{dbzeroinegr}) means,
we define a (Neumann) projector $\pi^\mu_{\,\,\nu} \equiv
\delta^\mu_{\,\,\nu}-Q^\mu_{\,\,\nu}$ with the properties
\beq
\pi^\mu_{\,\,\nu}\pi^\nu_{\,\,\rho}=\pi^\mu_{\,\,\rho},
\,\,\,\,\,\,\,\,\,\,\,\,\,\,\,
\pi^\mu_{\,\,\nu} Q^\nu_{\,\,\rho}= 0 .
\eeq{proppi}
We now have two orthogonal projectors $\pi$ and $Q$, corresponding
to the $P$ and $Q$ of the $B=0$ case. They may be written in terms of
a (1,1) ``tensor'' $r^\mu_{\,\,\nu}$ that squares to one, in analogy with
the ``tensor'' $R^\mu_{\,\,\nu}$ for $B=0$:
\beq
\pi^\mu_{\,\,\nu} = \half \left( \delta^\mu_{\,\,\nu}
+ r^\mu_{\,\,\nu} \right),
\,\,\,\,\,\,\,\,\,\,\,\,\,\,\,
Q^\mu_{\,\,\nu} = \half \left( \delta^\mu_{\,\,\nu}
- r^\mu_{\,\,\nu} \right) ,
\,\,\,\,\,\,\,\,\,\,\,\,\,\,\,
r^\mu_{\,\,\rho} r^\rho_{\,\,\nu}=\delta^\mu_{\,\,\nu} .
\eeq{e:piQdef}
Among the many useful relations that follow from all the
definitions above, there is the property that $\pi$
leaves $P$ invariant,
\beq
\pi^\mu_{\,\,\nu} P^\nu_{\,\,\rho}= P^\mu_{\,\,\rho} ,
\eeq{e:Pinvariant}
which is due to $Q^\mu_{\,\,\nu}P^\nu_{\,\,\rho}=0$.
Also note that the two boundary conditions (\ref{alferm}) and (\ref{bbcNt})
may be written in terms of 1D superfields analogously to what we did
in Appendix~B in reference \cite{Albertsson:2001dv}:
$DK^\mu = \pi^\mu_{\,\,\nu} (K) DK^\nu$ and
$S^\mu= Q^\mu_{\,\,\nu}(K) S^\nu$.

Returning now to condition (\ref{dbzeroinegr}), we may
use (\ref{e:Pinvariant}) to rewrite it as
\beq
P^\gamma_{\,\,\nu} P^\phi_{\,\,\sigma} \pi^\mu_{\,\,\gamma}
\pi^\rho_{\,\,\phi} \nabla_{[\rho} Q^\delta_{\,\,\mu]} = 0,
\eeq{pppipi}
which implies that $\pi$ satisfies the integrability condition,
\beq
\pi^\mu_{\,\,\gamma} \pi^\rho_{\,\,\phi}
  \nabla_{[\rho} Q^\delta_{\,\,\mu]} = 0 ,
\eeq{piinegrabil}
since $P$ is invertible on the $\pi$-subspace. We thus have
a situation completely analogous to that of the $B=0$ case, in that
the Neumann projector must be integrable. This integrability
condition turns out to be essential in the geometrical interpretation of
the supersymmetric boundary conditions, as we will see in
Section~\ref{s:geometry}.

Note that the ansatz (\ref{alferm}) for the fermionic boundary
conditions is a rather simple one, $R^\mu_{\,\,\nu}$ depending only on
$X^\mu$ and not on $\psi_+^\mu$.  The relation of this linear ansatz
to more general boundary conditions is discussed in
Section~\ref{s:discussion}.

\Section{N=1 superconformal symmetry}
\label{s:currents}

In this section we derive the full set of boundary conditions using
conserved currents. The bulk action (\ref{action1}) yields the following
supercurrents (the main steps in deriving them
were sketched in Appendix~C in reference \cite{Albertsson:2001dv}),
\beq
 T_{\+}^- = D_{+}\Phi^\mu \partial_{\+} \Phi^\nu g_{\mu\nu} - \frac{i}{3}
 D_{+}\Phi^\mu D_{+} \Phi^\nu D_{+} \Phi^\rho H_{\mu\nu\rho} ,
\eeq{superc1}
\beq
 T_{=}^+ = D_{-}\Phi^\mu \partial_{=} \Phi^\nu g_{\mu\nu} + \frac{i}{3}
 D_{-}\Phi^\mu D_{-} \Phi^\nu D_{-} \Phi^\rho H_{\mu\nu\rho} ,
\eeq{superc2}
which obey the corresponding conservation laws,
\beq
 D_{+}T_{=}^+ =0,\,\,\,\,\,\,\,\,\,\,\,\,\,\,\, D_{-}T_{\+}^- =0 .
\eeq{conservlaws}
The components of the supercurrents (\ref{superc1}) and
(\ref{superc2}) correspond to the supersymmetry
current and stress tensor as follows,
\beq
G_{+} = T_{\+}^-| = \psi_{+}^\mu \d_{\+} X^\nu g_{\mu\nu} - \frac{i}{3}
\psi_{+}^\mu \psi_{+}^\nu \psi_{+}^\rho H_{\mu\nu\rho} ,
\eeq{comp1}
\beq
G_{-} = T_{=}^+|= \psi_{-}^\mu \d_{=} X^\nu g_{\mu\nu} + \frac{i}{3}
\psi_{-}^\mu \psi_{-}^\nu \psi_{-}^\rho H_{\mu\nu\rho} ,
\eeq{comp2}
\beq
T_{++} = -iD_{+} T_{\+}^-| = \d_{\+}X^\mu \d_{\+}X^\nu g_{\mu\nu} + i
\psi^\mu_+ \nabla^{(+)}_{+} \psi^\nu_{+} g_{\mu\nu} ,
\eeq{comp3}
\beq
T_{--} = - iD_{-} T_{=}^+| =  \d_{=}X^\mu \d_{=}X^\nu g_{\mu\nu} +
 i \psi^\mu_- \nabla^{(-)}_{-} \psi^\nu_{-} g_{\mu\nu} ,
\eeq{comp4}
where the covariant derivatives acting on
the worldsheet fermions are defined by
\beq
\nabla^{(+)}_{\pm}\psi_{+}^\nu = \partial_{\pp}\psi_{+}^\nu +
\Gamma^{+\nu}_{\,\,\rho\sigma}\d_{\pp} X^\rho
\psi_{+}^\sigma,\,\,\,\,\,\,\,\,\,\, \nabla^{(-)}_{\pm}\psi_{-}^\nu =
\partial_{\pp}\psi_{-}^\nu + \Gamma^{-\nu}_{\,\,\rho\sigma}\d_{\pp}
X^\rho \psi_{-}^\sigma .
\eeq{covdervferm}

To ensure superconformal symmetry on the boundary
we need to impose the following boundary conditions on the
currents (\ref{comp1})--(\ref{comp4}),
\beq
 G_{+}-\eta G_{-} =0,\,\,\,\,\,\,\,\,\,\,\,\,\,\,\,\,\,
 T_{++}-T_{--}=0.
\eeq{superconf}
Classically these conditions make sense only on-shell, which
means that we may (and should) make use of the field equations in our
analysis. Thus we use the fermionic equations of motion,
\beq
g_{\mu\nu} (\psi_{+}^\mu \nabla^{(+)}_- \psi_+^\nu - \psi_{-}^\mu \nabla^{(-)}_+
\psi_-^\nu) = 0,
\eeq{fermeqdbo}
to rewrite the stress tensor condition as
\ber
\nonumber 0= T_{++}-T_{--} &=& \d_\+ X^\mu \d_\+ X^\nu g_{\mu\nu} - \d_=
X^\mu \d_= X^\nu g_{\mu\nu} + \\
\nonumber & +& i (\psi_{+}^\mu - \eta\psi_-^\mu) \nabla_0
(\psi_+^\nu + \eta \psi_-^\nu) g_{\mu\nu} + \\
\nonumber & +& i (\psi_+^\mu + \eta \psi_-^\mu)\nabla_0
(\psi_+^\nu-\eta \psi_-^\nu) g_{\mu\nu} + \\
&+& 2i(\psi^\mu_+ \psi_+^\nu 
 + \psi^\mu_- \psi_-^\nu )\d_0 X^\rho H_{\mu\rho\nu} ,
\eer{TTTnew}
where $\nabla_0$ is the covariant $\tau$-derivative
without torsion,
\beq
\nabla_0\psi_{\pm}^\nu = \partial_0\psi_{\pm}^\nu +
\Gamma^{\nu}_{\,\,\rho\sigma}\d_0 X^\rho \psi_{\pm}^\sigma .
\eeq{torsionfree}
Substituting the fermionic ansatz (\ref{alferm}) in
(\ref{TTTnew}) we get
\ber
\nonumber
0=T_{++} - T_{--} &=& 2i \psi_{+}^\sigma\d_0\psi_{+}^\lambda \left[
g_{\sigma\lambda} - R^\mu_{\,\,\sigma} g_{\mu\nu} R^\nu_{\,\,\lambda}
 \right] + \\
\nonumber & +& 2\d_0 X^\delta \pi^\rho_{\,\,\delta} \left[ g_{\rho\nu}
(\d_\+ X^\nu - \d_= X^\nu) - 2 B_{\rho\nu} \pi^\nu_{\,\,\lambda} \d_0
X^\lambda  +\right. \\
\nonumber  &+& \left. i \left( R^\mu_{\,\,\gamma}\Gamma_{\mu\rho\nu}
 R^\nu_{\,\,\sigma} -\Gamma_{\gamma\rho\sigma}
 - R^\mu_{\,\,\sigma} g_{\mu\nu} 
R^\nu_{\,\,\gamma,\rho} + \right. \right. \\
&+& \left. \left. H_{\sigma\rho\gamma} + R^\mu_{\,\,\sigma}
 H_{\mu\rho\nu} R^\nu_{\,\,\gamma} \right) \, \psi^\sigma_+ \psi^\gamma_+ 
   \right] .
\eer{TTTterms}
The extra term involving an antisymmetric field $B_{\mu\nu}$ and $\d_0 X^\lambda$
represents the arbitrariness due to contraction with $\d_0 X^\delta$
(the factors in front of $B_{\mu\nu}$ are fixed for convenience) and we add it
to find the most general boundary conditions.  Note that $B_{\mu\nu}$
in this term is a priori not necessarily the same as the background
$B$-field (i.e., the $B$-field whose field strength is
$H_{\mu\nu\rho}$); we just need a general antisymmetric field.
However, our choice is justified by the fact that our physical setup
provides the $B$-field as the only available antisymmetric
field. Moreover, supersymmetry will impose a relation between this
field and $H_{\mu\nu\rho}$ (see
Section~\ref{s:Bfield}) which strongly suggests that the extra term
in (\ref{TTTterms}) indeed contains $B_{\mu\nu}$.

Requiring that the first term in (\ref{TTTterms}) vanish independently
we obtain the following condition,
\beq
R^\mu_{\,\,\sigma} g_{\mu\nu} R^\nu_{\,\,\rho} =
g_{\sigma\rho} .
\eeq{gprop}
This condition, that $R^\mu_{\,\,\nu}$ preserves the metric, is our
first condition on $R^\mu_{\,\,\nu}$. It arose also for $B=0$, so we
see that turning on the background $B$-field does not change this
requirement.

We now substitute (\ref{gprop}) into (\ref{TTTterms}),
and find that the stress tensor condition reduces to
\beq
\left \{
\begin{array}{l}
 \pi^\rho_{\,\,\delta} E_{\nu\rho} \pi^\nu_{\,\,\lambda}
 \d_\+ X^\lambda -  \pi^\rho_{\,\,\delta} E_{\rho\nu} 
\pi^\nu_{\,\,\lambda} \d_= X^\lambda
- i\pi^\rho_{\,\,\delta} (R^\mu_{\,\,\sigma} g_{\mu\nu} \nabla_\rho
  R^\nu_{\,\,\gamma}  -  H_{\sigma\rho\gamma} - R^\mu_{\,\,\sigma}
 H_{\mu\rho\nu} R^\nu_{\,\,\gamma} )\psi^\sigma_+ \psi^\gamma_+ = 0, \\
 Q^\mu_{\,\,\lambda}(\d_= X^\lambda  + \d_\+ X^\lambda)= 0,
\end{array} \right .
\eeq{bosb111} 
where $E_{\mu\nu} \equiv g_{\mu\nu}+B_{\mu\nu}$, and the second equation
is just the assumption (\ref{QQ}). Moreover,
the condition on the supersymmetry current
in (\ref{superconf}) takes the form
(inserting the ansatz (\ref{alferm}) for $\psi_-^\mu$)
\beq
\begin{array}{ll}
0= G_+ - \eta G_- =& \psi_+^\sigma (g_{\sigma\nu} \d_\+ X^\nu -
R^\mu_{\,\,\sigma} g_{\mu\nu} \d_= X^\nu) \\
&- \frac{i}{3} \left( H_{\mu\nu\rho} + 
R^\sigma_{\,\,\mu} R^\lambda_{\,\,\nu} R^\gamma_{\,\,\rho}
 H_{\sigma\lambda\gamma} \right)
\psi_{+}^\mu \psi_{+}^\nu \psi_{+}^\rho .
\end{array}
\eeq{ggcondH}
Using (\ref{bosb111}) in (\ref{ggcondH})\footnote{Eq.~(\ref{ggcondH}) is
first rewritten using $\delta^\mu_{\,\,\nu}= \pi^\mu_{\,\,\nu} +
Q^\mu_{\,\,\nu}$ and $Q^\mu_{\,\,\nu} \d_0 X^\nu =0$.} and requiring
the first term to vanish we find
\beq
 \pi^\rho_{\,\,\delta} E_{\nu\rho} \pi^\nu_{\,\,\gamma} = \pi^\rho_{\,\,\delta}
 E_{\rho\nu} \pi^\nu_{\,\,\lambda} R^\lambda_{\,\,\gamma} .
\eeq{reqonR}
This is our second condition on
$R^\mu_{\,\,\nu}$.\footnote{In the case
of a space-filling D-brane, i.e., when there are no Dirichlet
directions, this condition amounts to $R =E^{-1}
E^t$, where $t$ denotes transposition (see Appendix~\ref{a:coords}
for a discussion of purely Neumann boundary conditions).}
  It turns out to be very useful in the continued
analysis; in particular, in our geometric discussion in
Section~\ref{s:geometry} we will use that the fermionic boundary
conditions (\ref{alferm}) take the form
\beq
\left \{
\begin{array}{l}
  \pi^\rho_{\,\,\delta} E_{\rho\nu} \pi^\nu_{\,\,\lambda}
 \psi_-^\lambda - \eta \pi^\rho_{\,\,\delta}
 E_{\nu\rho} \pi^\nu_{\,\,\gamma} \psi_+^\gamma = 0, \\
 Q^\mu_{\,\,\lambda}(\psi_-^\lambda  + \eta \psi_+^\lambda)= 0 .
\end{array} \right .
\eeq{fermbouncond}

Our third boundary condition is obtained by using (\ref{reqonR}) to
rewrite the remaining three-fermion term in (\ref{ggcondH}) as
\beq
\left [ P^\rho_{\,\,\tau} R^\mu_{\,\,\sigma} g_{\mu\nu} \nabla_\rho
  R^\nu_{\,\,\gamma} + \frac{4}{3} P^\mu_{\,\,\tau} P^\nu_{\,\,\sigma}
  P^\rho_{\,\,\gamma} H_{\mu\nu\rho} \right ] \psi_+^\tau
\psi_+^\sigma \psi_+^\gamma = 0 ,
\eeq{threeferm}
which implies that
\beq
P^\rho_{\,\,\tau} R^\mu_{\,\,\sigma} g_{\mu\nu} \nabla_\rho
R^\nu_{\,\,\gamma} + P^\rho_{\,\,\sigma} R^\mu_{\,\,\gamma} g_{\mu\nu}
\nabla_\rho R^\nu_{\,\,\tau}+ P^\rho_{\,\,\gamma} R^\mu_{\,\,\tau}
g_{\mu\nu} \nabla_\rho R^\nu_{\,\,\sigma} + 4 P^\mu_{\,\,\tau}
P^\nu_{\,\,\sigma} P^\rho_{\,\,\gamma} H_{\mu\nu\rho} = 0 .
\eeq{condsolved}
Contracting (\ref{condsolved}) with
$Q^\delta_{\,\,\tau}$ one arrives at the condition
\beq
P^\gamma_{\,\,\nu} P^\phi_{\,\,\sigma} \pi^\mu_{\,\,\gamma}
\pi^\rho_{\,\,\phi} \nabla_{[\rho} Q^\delta_{\,\,\mu]} = 0,
\eeq{pppipi11}
which is equivalent to integrability of $\pi^\mu_{\,\,\nu}$,
as we saw in Section~\ref{s:axioms}.

We summarise the conditions on $R^\mu_{\,\,\nu}$ we have obtained:
\beq
\left \{
\begin{array}{l}
\pi^\rho_{\,\,\delta} E_{\nu\rho} \pi^\nu_{\,\,\gamma} = \pi^\rho_{\,\,\delta}
 E_{\rho\nu} \pi^\nu_{\,\,\lambda} R^\lambda_{\,\,\gamma} , \\
\pi^\mu_{\,\,\gamma} \pi^\rho_{\,\,\nu}
\nabla_{[\rho} Q^\delta_{\,\,\mu]} = 0 , \\
 R^\mu_{\,\,\rho} g_{\mu\nu} R^\nu_{\,\,\sigma} = g_{\rho\sigma} .
\end{array} \right . 
\eeq{e:bcsum1}

\subsection{The B-field}
\label{s:Bfield}

We next investigate whether there are further conditions to be found,
by returning to the condition (\ref{threeferm}) and applying our newly
found conditions (\ref{e:bcsum1}).  The goal is to rewrite
(\ref{threeferm}) entirely in terms of covariant derivatives of
$B_{\mu\nu}$, to see what it says about the torsion.  To do this, we
need some relations between derivatives of $R^\mu_{\,\,\nu}$ and
derivatives of $B_{\mu\nu}$, and such relations are provided by
(\ref{e:bcsum1}) as follows.

A general $B$-field may be expanded in its $\pi$- and $Q$-constituents
as
$$
B_{\mu\nu} = B^{\cal D}_{\mu\nu} + B^{\cal O}_{\mu\nu},
$$
where the ``diagonal'' (${\cal D}$) and ``off-diagonal'' (${\cal O}$)
parts are defined as
\begin{eqnarray}
B^{\cal D}_{\mu\nu} & \equiv &
\pi^\sigma_{\,\,\mu} B_{\sigma\rho} \pi^\rho_{\,\,\nu}
+  Q^\sigma_{\,\,\mu} B_{\sigma\rho} Q^\rho_{\,\,\nu} ,
\label{e:BDdef} \\
B^{\cal O}_{\mu\nu} & \equiv &
\pi^\sigma_{\,\,\mu} B_{\sigma\rho} Q^\rho_{\,\,\nu}
+  Q^\sigma_{\,\,\mu} B_{\sigma\rho} \pi^\rho_{\,\,\nu}.
\label{e:BOdef}
\end{eqnarray}
Inserting (\ref{e:BDdef}) and (\ref{e:BOdef})
into (\ref{reqonR}) we obtain the relation
\beq
g_{\mu\nu} \left( R^\nu_{\,\,\gamma} - r^\nu_{\,\,\gamma} \right)
=-2 B^{\cal D}_{\mu\nu} P^\nu_{\,\,\gamma} ,
\eeq{e:gtheta}
the covariant derivative of which is
\beq
g_{\mu\nu} \left( \nabla_\rho R^\nu_{\,\,\gamma}
- \nabla_\rho r^\nu_{\,\,\gamma} \right)
=-2 P^\nu_{\,\,\gamma} \nabla_\rho B^{\cal D}_{\mu\nu}
- B^{\cal D}_{\mu\nu} \nabla_\rho R^\nu_{\,\,\gamma}.
\eeq{e:gthetaderiv}
Moreover, using (\ref{reqonR}) one may write
\beq
R^\mu_{\,\,\sigma} B_{\mu\nu} R^\nu_{\,\,\rho}=
B^{\cal D}_{\sigma\rho} +
R^\mu_{\,\,\sigma} B^{\cal O}_{\mu\nu} R^\nu_{\,\,\rho} ,
\eeq{e:RBR}
which is equivalent to
\beq
R^\mu_{\,\,\tau} B^{\cal D}_{\mu\nu} R^\nu_{\,\,\gamma}=
B^{\cal D}_{\tau\gamma} .
\eeq{e:RBDR}
Acting on this equation with $P^\rho_{\,\,\sigma} \nabla_\rho$
yields
\beq
P^\rho_{\,\,\sigma} B^{\cal D}_{\mu\nu} 
R^\mu_{\,\,[\tau} \nabla_{|\rho |} R^\nu_{\,\,\gamma]}
= P^\rho_{\,\,\sigma} \nabla_\rho B^{\cal D}_{\tau\gamma}
- P^\rho_{\,\,\sigma} R^\mu_{\,\,\tau} R^\nu_{\,\,\gamma}
\nabla_\rho B^{\cal D}_{\mu\nu} .
\eeq{e:PnablaRBDR}

Thus we have the relations we were looking for, namely
(\ref{e:gthetaderiv}) and (\ref{e:PnablaRBDR}),
and we use them, together with the definition
$P^\mu_{\,\,\nu}= \half (\delta^\mu_{\,\,\nu}+R^\mu_{\,\,\nu})$,
to rewrite (\ref{threeferm}) as
\beq
\psi_+^\sigma \psi_+^\tau \psi_+^\gamma \left[
2 P^\rho_{\,\,\sigma} P^\mu_{\,\,\tau} P^\nu_{\,\,\gamma}
\nabla_\rho B^{\cal D}_{\nu\mu} +
\frac{4}{3} P^\mu_{\,\,\tau} P^\nu_{\,\,\sigma}
  P^\rho_{\,\,\gamma} H_{\rho\mu\nu}
\right] =0 .
\eeq{e:PPPnablaBD}
This implies that
\beq
\pi^\mu_{\,\,\tau} \pi^\nu_{\,\,\sigma} \pi^\rho_{\,\,\gamma}
H_{\mu\nu\rho} = \frac{1}{2} \pi^\mu_{\,\,\tau} \pi^\nu_{\,\,\sigma}
\pi^\rho_{\,\,\gamma} (\nabla_\mu B^{\cal D}_{\nu\rho} + \nabla_\nu
B^{\cal D}_{\rho\mu}
+ \nabla_\rho B^{\cal D}_{\mu\nu}) .
\eeq{e:PPPH}
This condition gives us information about the way in
which our ad hoc introduced antisymmetric tensor $B_{\mu\nu}$ (see
Eq.~(\ref{TTTterms})) is related to the torsion $H_{\mu\nu\rho}$. We see that
the fully $\pi$-projected torsion equals
the corresponding part of the torsion of the
``diagonal'' part of the introduced tensor $B_{\mu\nu}$.
This lends strong support to our earlier assumption that the
introduced tensor in (\ref{TTTterms}) is in fact the background $B$-field.
Assuming that this is indeed the case, (\ref{e:PPPH}) reduces to
\beq
\pi^\mu_{\,\,\tau} \pi^\nu_{\,\,\sigma}
\pi^\rho_{\,\,\gamma} (\nabla_\mu B^{\cal O}_{\nu\rho} + \nabla_\nu
B^{\cal O}_{\rho\mu}
+ \nabla_\rho B^{\cal O}_{\mu\nu}) = 0 ,
\eeq{e:pipipiBO}
which is automatically satisfied since $\pi$ is integrable.
Thus we find no further conditions in addition to (\ref{e:bcsum1}).

As an aside one may consider some special cases of D-branes, to see
what happens to the relations derived above. For example, take
$B_{\mu\nu}=B^{\cal O}_{\mu\nu}$, i.e., the diagonal part is zero.
Then it follows immediately from (\ref{e:gtheta}) that
$R^\mu_{\,\,\nu} =r^\mu_{\,\,\nu}$. As a consequence
(because $\pi^\mu_{\,\,\rho} r^\rho_{\,\,\nu} =
 \pi^\mu_{\,\,\nu}$ and $Q^\mu_{\,\,\rho} r^\rho_{\,\,\nu} =
- Q^\mu_{\,\,\nu}$) the definition of $B^{\cal
O}_{\mu\nu}$
implies that
\beq
R^\mu_{\,\,\sigma} B_{\mu\nu} R^\nu_{\,\,\rho}=
-B_{\sigma\rho}.
\eeq{e:RBR_B}
This D-brane is oriented in such a way that effectively it does not feel
the $B$-field.
In addition, it follows trivially from (\ref{e:pipipiBO}) that
\beq
\pi^\mu_{\,\,\tau} \pi^\nu_{\,\,\sigma} \pi^\rho_{\,\,\gamma} H_{\mu\nu\rho}
 = 0 ,
\eeq{e:PPPHis0}
i.e., the fully $\pi$-projected torsion vanishes.

On the other hand, if $B_{\mu\nu}=B^{\cal D}_{\mu\nu}$, 
then (\ref{e:RBDR}) yields
\beq
R^\mu_{\,\,\sigma} B_{\mu\nu} R^\nu_{\,\,\rho}=
B_{\sigma\rho} .
\eeq{e:RBRB}
In this case, however, there is no additional information about the
torsion.

If $B$ is a symplectic form on ${\cal M}$ (i.e., $\det B \neq
0$ and $dB=0$), then the solution (\ref{e:RBR_B}) corresponds to a
Lagrangian submanifold of ${\cal M}$, and
(\ref{e:RBRB}) corresponds to a symplectic submanifold.

\subsection{Compatibility with the algebra}
\label{s:compalg}

In the interest of consistency, the previously derived results should
be compatible with the supersymmetry algebra (\ref{compsusytr}).  This
is verified by showing that the supersymmetry partner
(\ref{susybosRRR}) of the fermionic ansatz (\ref{alferm}) is
equivalent to the bosonic boundary condition (\ref{bosb111}) when the
conditions on $R$ are imposed. We do this by
using  the properties (\ref{QR}) of $Q$,
together with the ansatz (\ref{alferm}) and the  conditions
(\ref{e:bcsum1}), to rewrite (\ref{bosb111}) as
\beq
\d_= X^\mu - R^\mu_{\,\,\nu}\d_\+ X^\nu + 2i(P^\sigma_{\,\,\rho}
\nabla_\sigma R^\mu_{\,\,\nu} + P^\mu_{\,\,\gamma} g^{\gamma\delta}
H_{\delta\sigma\rho} R^\sigma_{\,\,\nu})\psi_+^\rho \psi_+^\nu =0 ,
\eeq{susybosRRR2}
which is precisely (\ref{susybosRRR}). Thus
we may safely conclude that the derived boundary conditions
are indeed compatible with the
supersymmetry algebra (\ref{compsusytr}).

\Section{The N=1 ${\bf \sigma}$-model action revised}
\label{s:action1}

It would be nice to rederive the results of Section~\ref{s:currents}
from a different approach, e.g., from an action.
However, in the presence of a $B$-field there is a
problem with the standard superfield action (\ref{action1}).
In this section we explain this problem and show how to cure it.  This
is in effect a clarification and extension of the
discussion in \cite{Haggi-Mani:2000uc}.

We start from the action (\ref{action1}) and for the sake of
simplicity we take $E_{\mu\nu}$ to be constant. The field variation
of the action produces a boundary term
\beq
\delta S = -i \int d\tau \,\,\left [ D_- (\delta \Phi^\mu D_- \Phi^\nu
E_{\mu\nu}) - D_+ (D_+ \Phi^\mu \delta \Phi^\nu E_{\mu\nu}) \right ] ,
\eeq{2S}
or in components,
\beq
\delta S = -i \int d\tau \,\,\left [ i\delta X^\mu (E_{\mu\nu} \d_=
X^\nu - E_{\nu\mu} \d_\+ X^\nu) + (\delta \psi_-^\mu \psi_-^\nu
E_{\mu\nu} - \delta \psi_+^\mu \psi_+^\nu E_{\nu\mu})\right ] .
\eeq{3S}
Assuming that we are interested in the free open string (i.e.,
$\delta X^\mu$ is arbitrary), we see from (\ref{3S}) that the bosonic
boundary condition should be
\beq
E_{\mu\nu} \partial_= X^\nu -
 E_{\nu\mu} \partial_\+ X^\nu = 0 .
\eeq{4S}
Now applying the supersymmetry transformation (\ref{compsusytr})
to (\ref{4S}), we obtain the fermionic condition
\beq
 E_{\mu\nu} \psi_-^\nu -
 \eta E_{\nu\mu} \psi_+^\nu = 0 ,
\eeq{5S}
where we have assumed that the left- and rightmoving supersymmetry
parameters are related as $\epsilon^+ =\eta \epsilon^-$. However,
inserting the two conditions (\ref{4S}) and (\ref{5S})
into (\ref{3S}) gives us a nonzero variation,
\beq
\delta S = -i \int d\tau \,\, 2 B_{\mu\nu}\delta \psi_-^\mu \psi_-^\nu .
\eeq{6S}  
(Alternatively one could write the integrand as $2 B_{\mu\nu}\delta
\psi_+^\mu \psi_+^\nu$ or $ B_{\mu\nu}\delta \psi_+^\mu \psi_+^\nu +
B_{\mu\nu}\delta \psi_-^\mu \psi_-^\nu$.) We conclude that the
supersymmetry algebra is incompatible with the requirement that
(\ref{3S}) vanish.

The problem with the action (\ref{action1}) is thus that it does
not allow supersymmetric freely moving open strings.  The most obvious
way to cure the problem is to add to the action
an extra boundary term, to compensate
for (\ref{6S}), e.g., $i \int d\tau \, B_{\mu\nu}\psi_-^\mu \psi_-^\nu$
or $i \int d\tau \, B_{\mu\nu}\psi_+^\mu \psi_+^\nu$.  In
\cite{Haggi-Mani:2000uc} the latter term was added, and it was shown
that the new action does admit supersymmetric freely moving open
strings.

However, if one tries to analyse the problem in all generality, i.e.,
to find an action which produces all solutions that we obtained from
the analysis of currents, then the boundary term $i \int d\tau \,
B_{\mu\nu}\psi_+^\mu \psi_+^\nu$ is not the right one. After some
trial and error one realises that the correct boundary term is $
\frac{i}{2} \int d\tau \,\, B_{\mu\nu} (\psi_+^\mu \psi_+^\nu
+\psi_-^\mu \psi_-^\nu )$, so that the modified action is
\beq
S= \int d^2\xi d^2\theta\,\, D_{+} \Phi^\mu D_{-} \Phi^\nu E_{\mu\nu}
(\Phi) -\frac{i}{2}\int d^2\xi\,\,  \partial_{=}( B_{\mu\nu}\psi_{+}^\mu
\psi_{+}^\nu + B_{\mu\nu} \psi_-^\mu \psi_-^\nu) .
\eeq{action111}
In the following section we show that this action indeed reproduces
all the solutions previously obtained from the currents.  The extra
boundary term in (\ref{action111}) has also appeared in \cite{Bachas}.

We may have fixed the supersymmetry problem, but now
there is another puzzle associated with the new action,
concerning the coupling to a $U(1)$ field. If we
shift the $B$-form by an exact two-form $d\Lambda$, then
(\ref{action111}) transforms as
\beq
 S(B+d\Lambda) = S(B) + 2 \int d\tau\,\,\Lambda_\mu \d_0 X^\mu ,
\eeq{shiftaction} 
where we expect the $\Lambda$-term to be absorbed
by a $U(1)$ coupling, because there should be a shift symmetry
\beq
 B \rightarrow B + d\Lambda,\,\,\,\,\,\,\,\,\,\,\,\,\,\,\,\,\,\,\,\,
 A \rightarrow A - \Lambda .
\eeq{shiftsym}
However, the last term in (\ref{shiftaction}) is purely bosonic and 
differs from the standard supersymmetric $U(1)$ coupling
\beq
\int d\tau\,\,\left [ 2 A_\mu \d_0 X^\mu -i \d_{[\mu} A_{\nu]}
(\psi_+^\mu + \eta \psi_-^\mu) (\psi_+^\nu + \eta \psi_-^\nu) \right
] .
\eeq{U1coupling} 

As a final general remark,
using the formal rules for 2D superfields one can write the action
(\ref{action111}) entirely in terms of superfields, as follows,
\beq
  S = \int d^2\xi d^2\theta\,\,\left [ D_{+} \Phi^\mu D_{-} \Phi^\nu
E_{\mu\nu} +
 \frac{1}{2D_+}D_-(B_{\mu\nu}D_+ \Phi^\mu D_+ \Phi^\nu + B_{\mu\nu} D_- \Phi^\mu D_- \Phi^\nu   ) \right ].  
\eeq{sueprfielac}
The last ``non-local'' term appears only when $B\neq 0$.  This form of
the action is perhaps formally acceptable, but we lack a
physical interpretation for this kind of non-locality in
superspace.

\Section{Boundary conditions from the action}
\label{s:action2}

Our aim in this section is to rederive the full set of
boundary conditions  from the action
\beq
S= \int d^2\xi d^2\theta\,\, D_{+} \Phi^\mu D_{-} \Phi^\nu E_{\mu\nu}
(\Phi) -\frac{i}{2}\int d^2\xi\,\,  \partial_{=}( B_{\mu\nu}\psi_{+}^\mu
\psi_{+}^\nu + B_{\mu\nu} \psi_-^\mu \psi_-^\nu) ,
\eeq{goodaction}
where now $E_{\mu\nu}$ is general.
We will show that the result coincides with the conditions obtained
from the currents in Section~\ref{s:currents}, and hence that the
above action is the correct one for describing the boundary dynamics.

The boundary term in the field variation of $S$ is given by
\ber
\nonumber \delta S &=& i \int d\tau \left [ (\delta \psi_+^\mu \psi_+^\nu -
\delta \psi_-^\mu \psi_-^\nu)g_{\mu\nu} + \right . \\
&& \left .+ \delta X^\mu (i\d_\+ X^\nu
E_{\nu\mu} - i\d_= X^\nu E_{\mu\nu} + \Gamma^-_{\nu\mu\rho} \psi_-^\nu
\psi_-^\rho - \Gamma^+_{\nu\mu\rho} \psi_+^\nu \psi_+^\rho) \right ] .
\eer{vargood}
When we insert the fermionic ansatz (\ref{alferm}) and use
the properties (\ref{QR}), cancellation of the fermionic variation
requires that
\beq
 R^\mu_{\,\,\sigma} g_{\mu\nu} R^\nu_{\,\,\rho} = g_{\sigma\rho}.
\eeq{metricprop}
This is the first of our conditions,
preservation of the metric.

Using (\ref{metricprop}), the variation (\ref{vargood}) collapses to
\ber
\nonumber \delta S &=& \int d\tau\,\,\delta X^\mu \left [ \d_= X^\nu
E_{\mu\nu} - \d_\+ X^\nu E_{\nu\mu} - \right. \\
&& \left . -
i(R^\gamma_{\,\,\rho} g_{\gamma\sigma} \nabla_\mu R^\sigma_{\,\,\nu} +
H_{\nu\mu\rho} + H_{\gamma\mu\sigma} R^\gamma_{\,\,\nu}
R^\sigma_{\,\,\rho}) \psi_+^\nu \psi_+^\rho \right ] ,
\eer{newvar}
implying the following bosonic boundary condition,
\ber
\nonumber \d_= X^\nu \pi^\mu_{\,\,\delta}E_{\mu\nu} &-& \d_\+ X^\nu
\pi^\mu_{\,\,\delta}E_{\nu\mu} - \\
&-& i\pi^\mu_{\,\,\delta} \left(R^\gamma_{\,\,\rho} g_{\gamma\sigma} \nabla_\mu
R^\sigma_{\,\,\nu} + H_{\nu\mu\rho} + H_{\gamma\mu\sigma}
R^\gamma_{\,\,\nu} R^\sigma_{\,\,\rho} \right)\, \psi_+^\nu
\psi_+^\rho =0 ,
\eer{e:bosbc1}
where we have assumed that $Q^\mu_{\,\,\nu} \delta X^\nu=0$,
or equivalently,
\beq
Q^\mu_{\,\,\nu}\left(\d_\+X^\nu + \d_= X^\nu\right)=0 .
\eeq{e:bosbc2}
(Note
that the condition that (\ref{newvar}) be zero gives rise to the same
kind of subtlety as did the vanishing of (\ref{TTTterms}); the terms
involving $B_{\mu\nu}$ vanish due to contraction with $\delta
X^\mu$. However, here this arbitrariness is automatically represented
by $B_{\mu\nu}$, as a physical and logical consequence of the action,
and we need not introduce an ad hoc field.)
The properties (\ref{metricprop}) and (\ref{e:bosbc2}) may now be used
to rewrite (\ref{e:bosbc1}) as
\ber
\nonumber 
\d_= X^\nu \pi^\mu_{\,\,\delta}E_{\mu\lambda}\pi^\lambda_{\,\,\nu} &-& 
 \d_\+ X^\nu \pi^\mu_{\,\,\delta}E_{\lambda\mu} \pi^\lambda_{\,\,\nu} -
\\
& -& i\pi^\mu_{\,\,\delta}(R^\gamma_{\,\,\rho}
 g_{\gamma\sigma} \nabla_\mu R^\sigma_{\,\,\nu} +
 H_{\nu\mu\rho} + H_{\gamma\mu\sigma} R^\gamma_{\,\,\nu} R^\sigma_{\,\,\rho})
 \psi_+^\nu \psi_+^\rho =0 ,
\eer{newboscond}
which is identical to (\ref{bosb111}).

Requiring supersymmetry on the boundary means, in terms of the
action, that the boundary conditions on the worldsheet fields
must be such that the field and supersymmetry variations
vanish simultaneously. Our next step is therefore to
examine the supersymmetry variation of (\ref{goodaction}),
which is given by
\ber
\nonumber
\delta_s S &=& \epsilon^- \int d\tau \,\,\left [ \d_\+ X^\mu
\psi_-^\nu E_{\mu\nu} - \eta \psi_+^\mu \d_= X^\nu E_{\mu\nu} + \eta
\d_\+ X^\mu \psi_+^\nu B_{\mu\nu} + \d_= X^\mu \psi_-^\nu
B_{\mu\nu}-\right . \\
\nonumber &&- \left . \frac{i}{3}\eta H_{\mu\nu\rho}
\psi_+^\rho \psi_+^\mu \psi_+^\nu - \frac{i}{3} H_{\mu\nu\rho}
\psi_-^\rho \psi_-^\mu \psi_-^\nu + \right . \\
 &&+ \left . i
F_{+-}^\mu (\eta \psi_-^\nu + \psi_+^\nu) g_{\mu\nu} + i
\Gamma^-_{\nu\mu\rho} \psi_+^\mu \psi_-^\rho \psi_+^\nu + i
\Gamma^-_{\nu\mu\rho} \psi_+^\mu \psi_-^\rho \psi_-^\nu \right ] .
\eer{susyvaraction}
Inserting the $F$-equation (\ref{Feom}), (\ref{susyvaraction}) becomes
\ber
\nonumber
\delta_s S &=& \epsilon^- \int d\tau \,\,\left [ \d_\+ X^\mu \psi_-^\nu
E_{\mu\nu} - \eta \psi_+^\mu \d_= X^\nu E_{\mu\nu} + \eta \d_\+ X^\mu
\psi_+^\nu B_{\mu\nu} + \d_= X^\mu \psi_-^\nu B_{\mu\nu}-\right .
\\
 &&- \left . \frac{i}{3}\eta H_{\mu\nu\rho} \psi_+^\rho \psi_+^\mu
\psi_+^\nu - \frac{i}{3} H_{\mu\nu\rho} \psi_-^\rho \psi_-^\mu
\psi_-^\nu \right ] .
\eer{afterFeq}
Now we plug in the fermionic ansatz (\ref{alferm}),
the bosonic conditions (\ref{e:bosbc1}), and the
property (\ref{e:bosbc2}), into
(\ref{afterFeq}). The $\d X \psi$-terms that remain after
this operation cancel only if
\beq
  \pi^\rho_{\,\,\delta} E_{\nu\rho} \pi^\nu_{\,\,\gamma} =
\pi^\rho_{\,\,\delta} E_{\rho\nu} \pi^\nu_{\,\,\lambda}
R^\lambda_{\,\,\gamma} .
\eeq{EERR}
Imposing (\ref{EERR}) thus reduces the condition
$\delta_s S=0$ to the 
following requirement for the three-fermion term,
\beq
\left [ P^\rho_{\,\,\tau} R^\mu_{\,\,\sigma} g_{\mu\nu} \nabla_\rho
  R^\nu_{\,\,\gamma} + \frac{4}{3} P^\mu_{\,\,\tau} P^\nu_{\,\,\sigma}
  P^\rho_{\,\,\gamma} H_{\mu\nu\rho} \right ] \psi_+^\tau
\psi_+^\sigma \psi_+^\gamma = 0 ,
\eeq{threeftc}
which is precisely the condition (\ref{threeferm}).
We know from Section~\ref{s:currents} that this equation
produces the integrability condition for $\pi$ ,
\beq
 \pi^\mu_{\,\,\sigma} \pi^\nu_{\,\,\gamma} Q^\rho_{\,\,[\mu,\nu]} =0 ,
\eeq{icp}
as well as the requirement that
\beq
\pi^\mu_{\,\,\tau} \pi^\nu_{\,\,\sigma} \pi^\rho_{\,\,\gamma}
H_{\mu\nu\rho} = \frac{1}{2} \pi^\mu_{\,\,\tau} \pi^\nu_{\,\,\sigma}
\pi^\rho_{\,\,\gamma} (\nabla_\mu B^{\cal D}_{\nu\rho} + \nabla_\nu
B^{\cal D}_{\rho\mu}
+ \nabla_\rho B^{\cal D}_{\mu\nu}) .
\eeq{pppHbd}
In Section~\ref{s:Bfield}
this condition gave us information about the way in which the ad hoc
introduced antisymmetric tensor $B_{\mu\nu}$ is related to
$H_{\mu\nu\rho}$.  Within the present analysis, however, we already
know that $H=dB$ (by definition), and
therefore (\ref{pppHbd}) leads directly to
\beq
\pi^\mu_{\,\,\tau} \pi^\nu_{\,\,\sigma}
\pi^\rho_{\,\,\gamma} (\nabla_\mu B^{\cal O}_{\nu\rho} + \nabla_\nu
B^{\cal O}_{\rho\mu}
+ \nabla_\rho B^{\cal O}_{\mu\nu}) = 0 ,
\eeq{Boffdcond}
i.e., condition (\ref{e:pipipiBO}). Due to (\ref{icp}), this is identically
satisfied.

Thus we have obtained the same set of boundary conditions as from the
currents, namely (\ref{metricprop}), (\ref{EERR}) and (\ref{icp})
(cf.~Eq.~(\ref{e:bcsum1})).  The check that the above boundary
conditions are compatible with the supersymmetry algebra
(\ref{compsusytr}) is now identical to that of
Section~\ref{s:compalg}, showing equivalence between
(\ref{newboscond}) and (\ref{susybosRRR}).

\Section{Geometric interpretation}
\label{s:geometry}

In this section we discuss the geometrical interpretation of our
results.  Some background information on submanifolds of Riemannian
manifolds is given in Appendix~\ref{a:submflds}.

\subsection{D-branes as submanifolds}

Let us first summarise the formal results we have derived
in previous sections. We found that supersymmetry requires
the worldsheet fields to obey the boundary conditions
\beq
\left \{
\begin{array}{l}
 \psi^\mu_- - \eta R^\mu_{\,\,\nu} \psi_+^\nu =0 , \\
 \d_= X^\mu - R^\mu_{\,\,\nu} \d_\+ X^\nu +2i (P^\sigma_{\,\,\rho}
 \nabla_\sigma R^\mu_{\,\,\nu}
 + P^\mu_{\,\,\gamma} g^{\gamma\delta} H_{\delta\sigma\rho}
 R^\sigma_{\,\,\nu})\psi_+^\rho \psi_+^\nu =0 ,
\end{array}
\right .
\eeq{formalres} 
where $2 P^\mu_{\,\,\nu} = \delta^\mu_{\,\,\nu}+R^\mu_{\,\,\nu}$,
and $R^\mu_{\,\,\nu}$ satisfies
\beq
\left \{
\begin{array}{l}
 \pi^\rho_{\,\,\delta} E_{\nu\rho} \pi^\nu_{\,\,\gamma}
= \pi^\rho_{\,\,\delta}
 E_{\rho\nu} \pi^\nu_{\,\,\lambda} R^\lambda_{\,\,\gamma} , \\
 Q^\mu_{\,\,\rho} R^\rho_{\,\,\nu} = R^\mu_{\,\,\rho} Q^\rho_{\,\,\nu}
= - Q^\mu_{\,\,\nu} , \\
 R^\mu_{\,\,\rho} g_{\mu\nu} R^\nu_{\,\,\sigma} = g_{\rho\sigma} , \\
 \pi^\mu_{\,\,\rho} \pi^\nu_{\,\,\sigma} Q^\lambda_{\,\,[\mu,\nu]} =0 .
\end{array}
\right .
\eeq{propR}
At first sight these results look rather formal.
However, it turns out they have a simple geometrical interpretation.

In mathematical terms, the Dirichlet projector $Q^\mu_{\,\,\nu}(X)$ is
a differentiable distribution\footnote{We need to assume
differentiability to be able to do the calculations. However one should
keep in mind that one can construct such brane configurations where
this property is lost (e.g., a brane ending on a brane).} which
assigns to a point $X$ in the $d$-dimensional spacetime manifold
${\cal M}$ a $(d-p-1)$-dimensional subspace\footnote{We take
$\rank(Q)=d-p-1$.}  of the tangent space $T_X({\cal M})$.  This
subspace consists of all vectors $v^\mu(X) \in T_X({\cal M})$ such
that
\beq
 Q^\mu_{\,\,\nu}(X) v^\nu(X) = v^\mu(X) .
\eeq{Qdef} 
The complementary distribution is defined as
$\pi^\mu_{\,\,\nu}=\delta^\mu_{\,\,\nu}- Q^\mu_{\,\,\nu}$,
and assigns to $X$ a $(p+1)$-dimensional space
that consists of vectors $v^\mu(X) \in T_X({\cal M})$ such that
\beq
 \pi^\mu_{\,\,\nu}(X) v^\nu(X) = v^\mu(X).
\eeq{pidef}
Now we ask when the vector fields defined by (\ref{pidef}) span a
submanifold. To see the answer, note that
the Lie bracket of two vector fields $v$ and $w$
in $\pi$-space is
\beq
\{ v, w\}^\nu = \pi^\nu_{\,\,\sigma} \{ v, w\}^\sigma +
v^\rho w^\sigma \pi^\mu_{\,\,\sigma} \pi^\lambda_{\,\,\rho}
Q^\nu_{\,\,[\lambda,\mu]} .
\eeq{Frobth}
If the last term vanishes (i.e., if $\pi^\mu_{\,\,\nu}$ is
integrable), then the distribution $\pi^\mu_{\,\,\nu}$ is involutive,
and due to the classical theorem of Frobenius there is a unique
\emph{maximal integral submanifold} corresponding to
$\pi^\mu_{\,\,\nu}$.  This submanifold of ${\cal M}$ is the
worldvolume of a D$p$-brane.  We emphasise that worldsheet
supersymmetry plays a crucial role in this interpretation.  Also note
that the discussion is completely local and therefore directly
applicable to our boundary conditions; there is no need to extend our
objects $R^\mu_{\,\,\nu}$, $\pi^\mu_{\,\,\nu}$, etc., to be globally
defined.

Having established that supersymmetric boundary conditions define
D$p$-branes as submanifolds of the spacetime Riemann manifold ${\cal
M}$, we may proceed to identify some of the objects associated with
such a submanifold.  In particular, there is an \emph{induced metric},
an \emph{induced connection}, a \emph{second fundamental form}, and an
\emph{associated second fundamental form}.
The induced metric may be taken to be the $\pi$-projected
part of $g_{\mu\nu}$. To identify the other structures,
take two vector fields $v$ and $w$ in the $\pi$-space.
Denoting by $\nabla$ the connection on ${\cal M}$, we may write
\beq
v^\mu \nabla_\mu w^\nu =
\pi^\nu_{\,\,\rho} v^\mu \nabla_\mu w^\rho
 + Q^\nu_{\,\,\rho} v^\mu \nabla_\mu w^\rho,
\eeq{e:nablavdecomp}
where we have decomposed the derivative into its tangential
(to the $\pi$-space) and normal parts
by using $\delta^\mu_{\,\,\nu} = \pi^\mu_{\,\,\nu} + Q^\mu_{\,\,\nu}$.
The tangential component is the induced connection, and
the normal component is the second fundamental form
(the definitions are given in Appendix~\ref{a:submflds}).
The latter may be rewritten,
using that $\pi^\mu_{\,\,\nu}v^\nu = v^\mu$ and
$\pi^\mu_{\,\,\nu}w^\nu = w^\mu$, as
\beq
v^\delta w^\sigma \, {\cal B}^\lambda_{\,\,\delta\sigma}
\equiv -v^\delta w^\sigma \,\, \pi^\mu_{\,\,\delta} \pi^\nu_{\,\,\sigma}
\nabla_\mu Q^\lambda_{\,\,\nu}.
\eeq{secff}
Note that ${\cal B}^\lambda_{\,\,\delta\sigma}$ is symmetric
in indices $\delta$ and $\sigma$, as a second fundamental form must be,
because $\pi^\mu_{\,\,\nu}$ is integrable.

Performing the same decomposition for the derivative of
a vector field $u$ in the $Q$-space, we have
\beq
v^\mu \nabla_\mu u^\nu =
\pi^\nu_{\,\,\rho} v^\mu \nabla_\mu u^\rho
 + Q^\nu_{\,\,\rho} v^\mu \nabla_\mu u^\rho,
\eeq{e:nablawdecomp}
where $Q^\mu_{\,\,\nu} u^\nu = u^\mu$ and $v$ is still in
the $\pi$-space.
The associated second fundamental form is then defined as the
tangential part, which we can rewrite as
\beq
v^\delta u^\sigma \, {\cal A}^\lambda_{\,\,\sigma\delta}
 \equiv - v^\delta u^\sigma \,\, \pi^\mu_{\,\,\delta}
\pi^\lambda_{\,\,\nu} \nabla_\mu Q^\nu_{\,\,\sigma} .
\eeq{asff}

For the case $B_{\mu\nu}=0$ the bosonic boundary
conditions in (\ref{formalres}) can be expressed using the associated
second fundamental form, as
\beq
\d_= X^\mu - R^\mu_{\,\,\nu} \d_\+ X^\nu + 4i {\cal
  A}^\mu_{\,\,\nu\gamma} \psi_+^\gamma \psi_+^\nu =0 .
\eeq{bosbzero}
Thus the properties of the two-fermion term are closely related to the
properties of the second fundamental form. For instance, when the
submanifold is totally geodesic (i.e., ${\cal A}={\cal B}=0$), the
two-fermion term vanishes and the bosonic boundary conditions
reduce to
\beq
\d_= X^\mu - R^\mu_{\,\,\nu} \d_\+ X^\nu =0 ,
\eeq{bosAzero}
a commonly assumed condition in the literature (see, e.g.,
\cite{Ooguri:1996ck}).

\subsection{Effects of the B-field}

When $B_{\mu\nu} \neq 0$, the two-fermion term will, in addition to
the associated second fundamental form, involve structures depending on
the $B$-field.  To study the effects of this on the boundary
conditions we introduce a parameter $z$ which measures the size of the
$B$-field, and then we vary this parameter within its allowed range.  We
define $B_{\mu\nu} \equiv zb_{\mu\nu}$, where $b_{\mu\nu}$ is a general
antisymmetric field, and the physically relevant situation is when $z$
can take any real value, $z \in \bR$.  One may now view
$R(z)$ as a matrix-valued function of $z$ for a given
point $X$ in ${\cal M}$. The behaviour of this function is determined
by the first two of Eqs.~(\ref{propR}),
\beq
\left\{
\begin{array}{l}
\pi^\rho_{\,\,\delta} (g_{\rho\nu}- zb_{\rho\nu})
 \pi^\nu_{\,\,\gamma} = \pi^\rho_{\,\,\delta}
 (g_{\rho\nu} +z b_{\rho\nu}) \pi^\nu_{\,\,\lambda} 
R^\lambda_{\,\,\gamma} (z) , \\
  Q^\mu_{\,\,\rho} R^\rho_{\,\,\nu}(z) = R^\mu_{\,\,\rho}(z)
 Q^\rho_{\,\,\nu} = - Q^\mu_{\,\,\nu} .
\end{array} \right.
\eeq{defzR}
An immediate consequence of (\ref{defzR}) is the property
\beq
R^\mu_{\,\,\nu}(z) R^\nu_{\,\,\gamma}(-z) = \delta^\mu_{\,\,\gamma},
\eeq{propRRI}
showing that $R(-z)$ is the inverse of $R(z)$.

To simplify the analysis of $R(z)$, we extend $z$ to the complex
plane, $z \in \bC$.  Then (\ref{defzR}) implies that $R(z)$ is a
meromorphic function with poles given by the condition
$\det(\pi^\rho_{\,\,\delta} (g_{\rho\nu} +z b_{\rho\nu})
\pi^\nu_{\,\,\lambda}) = 0$ (the determinant is understood to be taken
in the $\pi$-space). It is clear from this condition that the poles
are finite in number and located away from the real axis if the metric
is positive definite.\footnote{If the metric has Minkowski signature,
there may be a pole on the real axis away from the origin, associated
with the critical ``electric'' field \cite{Seiberg:2000ms}. In this
case one cannot go continuously from $z=0$ to $z=\infty$ while keeping
$z$ real.}

For the purpose of studying the limits $z\ra 0$ and $z\ra\infty$, we
write $R(z)$ as a Taylor expansion around each limit. The first series
is one in positive powers of $z$ around zero,
\beq
R^\mu_{\,\,\nu}(z) = r^\mu_{\,\,\nu} + \sum\limits_{k=1}^{\infty}
R^{(k)\mu}_{\,\,\nu} z^k, \,\,\,\,\,\,\,\,\,\,\,\,|z| < |z_{min}|
\eeq{Rzposit}
where $r^\mu_{\,\,\nu} = R^\mu_{\,\,\nu}(0)$ is independent of $z$,
and $z_{\min}$ is the pole closest to the origin.  Due to
(\ref{propRRI}), $r$ squares to one, $r^\mu_{\,\,\nu}
r^\nu_{\,\,\rho}=\delta^\mu_{\,\,\rho}$, a property that we discussed
in Section~\ref{s:axioms}.

The expansion around infinity can similarly be written as a power series in
$1/z$,
\beq
 R^\mu_{\,\,\nu}(z) = \tilde{r}^\mu_{\,\,\nu} + \sum\limits_{k=1}^{\infty} 
 \tilde{R}^{(k)\mu}_{\,\,\nu} z^{-k},\,\,\,\,\,\,\,\,\,\,\,\,|z| > |z_{max}| 
\eeq{Rznegatz}
where $z_{\max}$ is the pole most distant from the origin.  Again we
see that the constant $\tilde{r}$ squares to one.

The discussion so far has been restricted to a given point $X$ in
${\cal M}$, but as one moves on ${\cal M}$, the poles of $R(z)$ move
in $\bC$, since their location is determined by the metric, by the
projector $\pi^\mu_{\,\,\nu}$ and by $b_{\mu\nu}$, all of which depend
on $X$.  In principle it is possible for a pole to move out to
infinity, rendering (\ref{Rznegatz}) useless.  Thus, note that
$\tilde{r}$ has a quite different status from that of $r$; it is not
necessarily well-defined in the same way that $r$ is.  In particular, it
may not be a continuous function of $X$. This is easily seen for the
special case when $\pi^\mu_{\,\,\rho} b_{\mu\nu} \pi^\nu_{\,\,\sigma}$
changes rank; a small perturbation of $b$ causes a jump in
$\tilde{r}$.

The expansion (\ref{Rzposit}) is well-defined and safe to use as it
stands for analysing the $z\ra 0$ limit; the only possible subtlety
would be when there is a pole at the origin, but that never happens.
 Taking $z\ra 0$ just means that
$B_{\mu\nu}=0$, reducing our boundary conditions to the ones
derived in \cite{Albertsson:2001dv}.

Using the expansion (\ref{Rznegatz}), however, requires some care.
First, one needs to make sure that $\tilde{r}$ is well-defined and
differentiable on the neighbourhood in ${\cal M}$ where we want to
perform the analysis; second, there must exist a region of
convergence.
 The first issue depends on the rank of $\pi^\mu_{\,\,\rho} b_{\mu\nu}
\pi^\nu_{\,\,\sigma}$, as indicated above. Thus as long as we keep to
a fixed rank, $\tilde{r}$ is differentiable. The second issue is
related to the location of the poles, which depends on the spacetime
coordinates.

Hence we restrict our attention to a situation where
$\pi^\mu_{\,\,\rho} b_{\mu\nu} \pi^\nu_{\,\,\sigma}$ has a fixed rank
and where $g_{\mu\nu}$, $b_{\mu\nu}$ and $\pi^\mu_{\,\,\nu}$ do not vary significantly in
some neighbourhood in ${\cal M}$.  Then the expansion (\ref{Rznegatz})
makes sense in this neighbourhood and can be used to study the $z
\rightarrow \infty$ limit of our boundary conditions.
We proceed by looking at the boundary conditions on the form
(\ref{bosb111}) and (\ref{fermbouncond}), in which we substitute the
definition $B_{\mu\nu} = zb_{\mu\nu}$ and the expansion
(\ref{Rznegatz}). Then as $z \rightarrow \infty$,
(\ref{fermbouncond}) collapses to fermionic Dirichlet conditions along
$\pi^\mu_{\,\,\rho} b_{\mu\nu}\pi^\nu_{\,\,\sigma}$, whereas the
bosonic conditions (\ref{bosb111}) deviate from Dirichlet ones by a
two-fermion term involving the field strength $H_{\mu\nu\rho}$.  This deviation
poses a problem for the physical interpretation; if we want to picture
the boundary conditions in terms of D-branes, the bosonic condition
must also be pure Dirichlet. Hence we demand that the two-fermion
vanish, obtaining
\beq
 \pi^\mu_{\,\,\delta} \left( H_{\mu\nu\rho} +
\tilde{r}^\sigma_{\,\,\nu} \tilde{r}^\lambda_{\,\,\rho}
H_{\mu\sigma\lambda} \right) =0
\eeq{hpi}
in the limit $z\rightarrow\infty$.

In conclusion we see that, provided that $H_{\mu\nu\rho}$ satisfies
(\ref{hpi}) there is a flow between $r$ and $\tilde{r}$ in the sense
that those boundary conditions along $B_{\mu\nu}$
  which are Neumann for zero $B$-field
flow to Dirichlet conditions for very large $B$-field.

\Section{Deformation of almost product structures}
\label{s:global}

In our previous paper \cite{Albertsson:2001dv} we pointed out that,
in the absence of a $B$-field, globally defined supersymmetric
boundary conditions lead to the appearance of a partially integrable
almost product manifold, and vice versa.
In the present general case, with arbitrary $B_{\mu\nu}$, we find an
interesting generalisation of the almost product structure which is
worth discussing.

Let ${\cal M}$ be a $d$-dimensional manifold with a $(1,1)$ tensor
$r^\mu_{\,\,\nu}$ such that, globally,
\beq
 r^\mu_{\,\,\nu} r^\nu_{\,\,\rho} = \delta^\mu_{\,\,\rho} .
\eeq{rdefin}
Then ${\cal M}$ is an almost product manifold \cite{Yano2} with almost product
structure $r^\mu_{\,\,\nu}$. Assuming that there is a $(1,1)$ tensor
$b^\mu_{\,\,\nu}$ such that $b^\mu_{\,\,\nu} = - b_\nu^{\,\,\mu}$, one
can define a whole set of new $(1,1)$ tensors $R^\mu_{\,\,\nu}(z)$
as follows,
\beq
\begin{array}{l}
\pi^\rho_{\,\,\delta} (\delta^\delta_{\,\,\nu}- z 
b^\delta_{\,\,\nu}) \pi^\nu_{\,\,\gamma} = \pi^\rho_{\,\,\delta}
 (\delta^\delta_{\,\,\nu} +z b^\delta_{\,\,\nu}) \pi^\nu_{\,\,\lambda}
 R^\lambda_{\,\,\gamma} (z) , \\
  Q^\mu_{\,\,\rho} R^\rho_{\,\,\nu}(z) = R^\mu_{\,\,\rho}(z) 
Q^\rho_{\,\,\nu} = - Q^\mu_{\,\,\nu} ,
\end{array}
\eeq{defzR2}
where $z \in \bR$, and $Q^\mu_{\,\,\nu}= \half (\delta^\mu_{\,\,\nu}-
r^\mu_{\,\,\nu})$ is a globally defined distribution. It follows
from (\ref{defzR2}) that $R^\mu_{\,\,\nu}(z)$ satisfies
\beq
 R^\mu_{\,\,\nu}(z) R^\nu_{\,\,\gamma}(-z) = \delta^\mu_{\,\,\gamma} .
\eeq{propRRI2}
At any given point in ${\cal M}$ one can go to adapted coordinates,
where $R^\mu_{\,\,\nu}(z)$ can be symbolically written as
\beq
R(z) =\left (
\begin{array}{cc}
 \frac{I- zb}{I+zb} & 0 \\
  0 & -I
\end{array}
\right ) ,
\eeq{matrR}
where $b=(b^n_{\,\,m})$ (i.e., $b$ is along the $\pi$-directions) and
the inverse of $(I+zb)$ is understood to be taken in the $\pi$-space.
Since $R^\mu_{\,\,\nu}(0)=r^\mu_{\,\,\nu}$, one may think of
$R^\mu_{\,\,\nu}(z)$ as a deformation of the almost product 
structure $r^\mu_{\,\,\nu}$.

Following the logic of Section~\ref{s:geometry} one can extend $z$ to
the complex plane and study its analytic behaviour. For a given point
$X$ in ${\cal M}$, $R^\mu_{\,\,\nu}(z)$ is then a meromorphic function
with a finite number of poles given by $\det(I+zb)=0$ (as before the
determinant is understood to be taken in the $\pi$-space). Again we
assume that the number of poles does not change (i.e., the rank of $b$
does not change) and that they do not move much in $\bC$ as $X$ moves
in ${\cal M}$, so that we can use the expansions (\ref{Rzposit}) and
(\ref{Rznegatz}).  Then there will be a flow between $r$ and
$\tilde{r}$, just like in the previous section, which in the
present context is a flow between two almost product structures.

\Section{Discussion}
\label{s:discussion}

Starting from the fermionic ansatz
\beq
 \psi^\mu_-= \eta R^\mu_{\,\,\nu}(X)\psi_+^\nu
\eeq{ferman} 
and assuming that the bosonic coordinate is confined to some region of
the tangent manifold of spacetime (i.e., we take $Q^\mu_{\,\,\nu}\d_0
X^\nu =0$), we have derived the supersymmetric boundary conditions for
the general $N=1$ non-linear sigma model.  The problem is analysed in
two different ways: by studying the conserved currents and by studying
requirements for invariance of the action. In essence we find that the
D-brane has to be a submanifold of the target space,
and that the $B$-field and torsion have significant
effects on the boundary conditions.

It is natural to ask to what extent the present results are general.
In principle one could start from the most generic 
 form of fermionic boundary conditions \cite{Alvarez,Borlaf:1996na},
\beq
 \psi_-^\mu = {\cal R}^\mu (X, \psi_+) = \eta R^\mu_{\,\,\nu}(X)\psi_+^\nu +
 \eta  R^\mu_{\,\,\nu\rho\sigma}(X) \psi_+^\nu \psi_+^\rho
\psi_+^\sigma + ... ,
\eeq{generalform}
where the dots stand for terms with five or more fermions.  After
plugging (\ref{generalform}) into the stress tensor condition
$T_{++}-T_{--}=0$ on the form (\ref{TTTnew}), one realises without
much effort that the three-fermion term in (\ref{generalform}) gives
rise to a four-fermion term in the bosonic boundary conditions.
However, the two-fermion term in the bosonic condition would stay
exactly the same as in (\ref{bosb111}). In fact, since we keep the
property that $Q^\mu_{\,\,\nu}\d_0 X^\nu =0$, the general problem can
be solved order by order in the fermions. Thus in the present paper we
have derived the general supersymmetric boundary conditions up to
two-fermion terms, and the proposed geometrical interpretation in
terms of maximal integral manifolds will still be valid for the
generic case.  However, it is unclear to us what could be the
geometrical interpretation of $R^\mu_{\,\,\nu\rho\sigma}$ and other
higher rank objects. It is to be expected that the four- and
higher-order fermion terms must be included in the bosonic boundary
condition in the quantum theory. This problem would be interesting to
study further.

In the light of the above discussion the two-fermion terms in the
bosonic boundary conditions are important for consistency with the
supersymmetry algebra (see also Section~\ref{s:axioms}).  However, the
boundary conditions usually adopted in the literature (see, e.g.,
\cite{Ooguri:1996ck,Hori:2000ck}) do not have two-fermion terms.  As
we have explained, the two-fermion terms can be absent but only in
very specific situations, for instance when the D-brane is a totally
geodesic submanifold. We find this point confusing and think that this
issue deserves further investigation. To resolve the problem one has
to generalise the present results to the $N=2$ supersymmetric sigma
model, as this is the model studied in the above references.

Another aspect of our work which calls for further study is the
issue of the correct action for the $N=1$ supersymmetric sigma model
with boundaries. In particular, the problem of the supersymmetric
$U(1)$ coupling and the shift symmetry needs to be solved.

\bigskip

\bigskip

{\bf Acknowledgments}: We are grateful to Martin Ro\v{c}ek, Peter van
Nieuwenhuizen, Andrea Cappelli, Massimo Bianchi and Augusto Sagnotti
for useful discussions. MZ would like to thank the ITP, Stockholm
University, where part of this work was carried out.  UL acknowledges
support in part by EU contract HPNR-CT-2000-0122 and by NFR grant
650-1998368.

\bigskip

\bigskip

{\bf Note added in Proof:}
Due to editorial problems at a different journal, the publication of this
article has been delayed considerably. There has thus been some subsequent
development along the lines described here. More precisely, the techniques
have been applied to N=1,2 WZW-models in  \cite{Lindstrom:2002vp} and
\cite{Albertsson:2003va}. In these articles we analyze the
restrictions on the
gluing conditions of the affine currents and relate them to our boundary
conditions. We reproduce known result from this geometric point of view and
generalize them. In particular we find that the gluing map between the left
and right affine currents may be generalized in a very specific way
allowing for non-constant Lie algebra automorphisms.

\appendix

\Section{$(1,1)$ supersymmetry}
\label{a:susy}

Throughout the paper we use $\mu,\nu,...$ as spacetime indices,
$(\+,=)$ as worldsheet indices (in lightcone coordinates $\xi^\pm
\equiv \tau \pm \sigma$, where $\tau$, $\sigma$ are the usual
worldsheet coordinates), and $(+,-)$ as two-dimensional spinor
indices.  We also use superspace conventions where the pair of spinor
coordinates are labelled $\th^{\pm}$, and the covariant derivatives
$D_\pm$ and supersymmetry generators $Q_\pm$ satisfy
\ber
D^2_+ &=&i\d_\+, \quad
D^2_- =i\d_= , \quad \{D_+,D_-\}=0 , \cr
Q_\pm &=& -D_\pm+2i\th^{\pm}\d_{\pp} ,
\eer{alg}
where $\d_{\pp}=\partial_0\pm\partial_1$
($\d_{0,1} \equiv \d_{\tau,\sigma}$).
In terms of the covariant derivatives, a supersymmetry transformation of
a
superfield
$\P$ is then given by
\ber
\delta \P &\equiv & (\e^+Q_++\e^-Q_-)\P \cr
&=& -(\e^+D_++\e^-D_-)\P
+2i(\e^+\th^+\d_\++\e^-\th^-\d_=)\P
\eer{tfs}
The components of a superfield $\P$ are defined via projections as
follows,
\ber
\P|\equiv X, \quad D_\pm\P| \equiv \p_\pm, \quad D_+D_-\P|\equiv F_{+-}
,
\eer{comp}
where a vertical bar denotes ``the $\th =0$ part of ''.
Thus, in
components, the $(1,1)$ supersymmetry transformations are given by
\beq
\left \{ \begin{array}{l}
 \delta X^\mu = - \epsilon^{+} \psi_+^\mu - \epsilon^- \psi_-^\mu \\
 \delta \psi_+^\mu =  -i\epsilon^+ \d_{\+}X^\mu - \epsilon^- F^\mu_{-+}\\
\delta \psi_-^\mu  = -i \epsilon^- \d_{=} X^\mu - \epsilon^+ F_{+-}^\mu \\
\delta F^\mu_{+-} = - i \epsilon^+ \d_{\+} \psi_-^\mu + i \epsilon^- \d_= \psi_+^\mu
\end{array} \right .
\eeq{compsusytr}

\Section{Affine connection with torsion}
\label{a:connections}

Here we collect our conventions for the affine connection with and
without
 torsion.  The covariant derivatives
 are defined as follows,
\beq
 \nabla_\rho^{(\pm)} R^\mu_{\,\,\nu} = R^\mu_{\,\,\nu,\rho} +
\Gamma^{\pm\mu}_{\,\,\rho\sigma}R^\sigma_{\,\,\nu}
 - \Gamma^{\pm\sigma}_{\,\,\rho\nu}R^\mu_{\,\,\sigma},
\eeq{covderdef}
\beq
 \nabla_\rho R^\mu_{\,\,\nu} = R^\mu_{\,\,\nu,\rho} +
\Gamma^{\mu}_{\,\,\rho\sigma}R^\sigma_{\,\,\nu}
 - \Gamma^{\sigma}_{\,\,\rho\nu}R^\mu_{\,\,\sigma},
\eeq{covnormal}
 where the comma stands for the partial derivative, so that
$R^\mu_{\,\,\nu,\rho}
\equiv\d_\rho R^\mu_{\,\,\nu}$.
 The functions $\Gamma$ are defined as
\beq
\begin{array}{l}
 \Gamma^{\pm\nu}_{\,\,\rho\sigma} = \Gamma^{\nu}_{\,\,\rho\sigma} \pm
 g^{\nu\mu} H_{\mu\rho\sigma},\\
\Gamma_{\nu\rho\mu}=g_{\nu\sigma} \Gamma^{\sigma}_{\,\,\,\,\,\rho\mu}
 =\frac{1}{2}( g_{\nu\mu,\rho} +g_{\rho\nu,\mu} - g_{\rho\mu,\nu}),
\end{array}
\eeq{Gammapm}
 with $H$ being the torsion three form,
\beq
H_{\mu\rho\sigma} = \frac{1}{2}
 (B_{\mu\rho,\sigma} + B_{\rho\sigma,\mu} + B_{\sigma\mu,\rho}).
\eeq{Hdefin}
It follows that $\Gamma^+_{\mu\nu\rho}=\Gamma^-_{\mu\rho\nu}$.

\Section{Boundary conditions in different coordinate systems}
\label{a:coords}

By going to a specific coordinate system, the boundary conditions can
sometimes be simplified considerably.  As an illustration of this idea
we discuss below the purely Neumann boundary conditions in different
coordinate systems. The generalisation to other cases is
straightforward.

The Neumann boundary  conditions (freely moving string),
in the presence of arbitrary metric and $B$-field, are
\beq
 \left \{ \begin{array}{l}
 E_{\mu\nu}\psi_{-}^\nu - \eta E_{\nu\mu}\psi_{+}^\nu = 0 , \\
 E_{\mu\nu} \partial_{=}X^\nu - E_{\nu\mu}\partial_{\+} X^\nu + i\eta
\psi_{-}^\nu \psi_{+}^\rho
 E_{\rho\nu,\mu} + i \psi_{-}^\nu \psi_{-}^\rho E_{\mu\nu,\rho}
 - i\psi_{+}^\nu
 \psi_{+}^\rho E_{\nu\mu,\rho}=0 ,
\end{array} \right .
\eeq{fullNbc}
where $E_{\mu\nu} \equiv g_{\mu\nu} + B_{\mu\nu}$.
 The bosonic conditions  can be rewritten as
\ber
\nonumber E_{\mu\nu} \partial_{=}X^\nu - E_{\nu\mu}\partial_{\+} X^\nu
&+& i\Gamma_{\nu\rho\mu}
 (\psi_{+}^\rho -\eta \psi_-^\rho)(\psi_{+}^\nu +\eta \psi_{-}^\nu) +
\\
&+& i\eta  \psi_{-}^\nu \psi_{+}^\rho
 B_{\rho\nu,\mu} + i (\psi_{-}^\nu \psi_{-}^\rho + \psi_{+}^\nu
 \psi_{+}^\rho) B_{\mu\nu,\rho} = 0 .
\eer{newfbc}
If we consider the case when the $B$-field is covariantly constant (i.e.,
$\nabla_\rho B_{\mu\nu}=0$ with $\nabla_\rho$ being the Levi-Civita
connection), then in normal coordinates the $\Gamma$-part and
derivatives of $B_{\mu\nu}$ vanish.  Thus in this case there is always a
coordinate system where the boundary conditions take the simple form
\beq
 \left \{ \begin{array}{l}
 E_{\mu\nu}\psi_{-}^\nu - \eta E_{\nu\mu}\psi_{+}^\nu = 0 , \\
 E_{\mu\nu} \partial_{=}X^\nu - E_{\nu\mu}\partial_{\+} X^\nu =0 .
\end{array} \right .
\eeq{simpleNbc}
If the $B$-form is closed ($dB=0$), then locally it can be written as an
exact form $B_{\mu\nu}=\partial_{[\mu} A_{\nu]}$, and (\ref{newfbc})
becomes
\ber
\nonumber E_{\mu\nu} \partial_{=}X^\nu - E_{\nu\mu}\partial_{\+} X^\nu &+&
i\Gamma_{\nu\rho\mu}
 (\psi_{+}^\rho -\eta \psi_-^\rho)(\psi_{+}^\nu +\eta \psi_{-}^\nu) + \\
&+& i (\psi_{+}^\nu +\eta \psi_{-}^\nu)
 (\psi_{+}^\rho +\eta \psi_-^\rho)\partial_\mu \partial_\rho A_\nu =0.
\eer{newbcwA}
 In this case there is always a coordinate system (Darboux-like
 coordinates) where we can get rid of the derivatives of $B$
(i.e., where the last term in (\ref{newbcwA}) vanishes).

\Section{Submanifolds of Riemannian manifolds}
\label{a:submflds}

In this appendix we summarise the relevant mathematical details on
submanifolds of Riemannian manifolds.  In our use of terminology we
closely follow \cite{Yano2}.

We first give the definition of a \emph{distribution} on a manifold (or
neighbourhood) ${\cal M}$.  A distribution $\pi$ of dimension $(p+1)$
on ${\cal M}$ is an assignment to each point $X \in {\cal M}$ of a
$(p+1)$-dimensional subspace $\pi_X$ of the tangent space $T_X({\cal
M})$.  The assignment can be done in different ways, for instance by
means of an appropriate projection operator.  A distribution $\pi$ is called
\emph{differentiable} if every point $X$ has a neighbourhood $U$ and $(p+1)$
differentiable vector fields, which form a basis of $\pi_Y$ at every
$Y \in U$.  Furthermore, $\pi$ is called \emph{involutive}
if for any two vector fields
$v_i$, $v_j$ $\in \pi_X$ their Lie bracket $\{ v_i, v_j \} \in \pi_X$
for all $X \in {\cal M}$.

A connected submanifold $D$ of ${\cal M}$ is called an \emph{integral
manifold} of $\pi$ if $f_*(T_X(D))=\pi_X$ for all $X \in D$, where $f$
is the embedding of $D$ into ${\cal M}$.  If there is no other
integral manifold of $\pi$ which contains $D$, then $D$ is called a
\emph{maximal integral manifold} of $\pi$.

{\bf Frobenius theorem:} Let $\pi$ be an involutive distribution on a
manifold ${\cal M}$. Then through every point $X \in {\cal M}$, there
passes a unique maximal integral manifold $D(X)$ of $\pi$.  Any
other integral manifold through $X$ is an open submanifold of $D(X)$.

If the manifold ${\cal M}$ is Riemannian, then various structures may
be induced on the submanifold $D$. For instance, $D$ is automatically
Riemannian. If one defines the Levi-Civita connection
$\nabla_v \equiv v^\mu \nabla_\mu$ on ${\cal M}$, and takes two
vector fields $v$ and $w$ in the tangent space $T(D)$ of $D$, then
the covariant derivative $\nabla_v w$ can be decomposed as
\beq
 \nabla_v w =
\hat{\nabla} _v w + {\cal B}(v,w),
\eeq{decomcov}
where $\hat{\nabla}
_v w$ is the tangential component (i.e., it is in $T(D)$) and ${\cal
B}(v,w)$ is the
normal component. One can show that $\hat{\nabla} _v$ can serve as
the induced connection on the submanifold $D$. ${\cal B}$ is called the
\emph{second fundamental form} of $D$. Sometimes it is also useful to introduce
the \emph{associated second fundamental form}, ${\cal A}$, which
is defined as follows.  Taking $u$ to be
a normal vector field on $D$ and $v$ a tangent vector field on $D$ we write
\beq
 \nabla_v u = -{\cal A}_u v + D_v u 
\eeq{secassff}
 where $-{\cal
A}_u v$ and $D_v u$ are, respectively, the tangential and the normal
components of $\nabla_v u$. Using the metric $g$ on ${\cal M}$ one can
prove the following simple identity,
\beq
 g( {\cal B}(v,w), u)= g({\cal A}_u v, w) .
\eeq{ABrel}
Eqs.~(\ref{decomcov}) and
(\ref{secassff}) are called the Gauss formula and the Weingarten
formula, respectively.

\end{document}